\documentclass[aps,prb,preprint,groupedaddress,superscriptaddress,showpacs]{revtex4}
\usepackage[pdftex]{graphicx}
\usepackage{amsmath}
\begin{document}

\title{Mesodynamics with implicit degrees of freedom}

\author{Keng-Hua Lin}
\affiliation{School of Materials Engineering and Birck Nanotechnology Center,
Purdue University, West Lafayette, IN 47907, USA}

\author{Brad Lee  Holian} 
\affiliation{Theoretical Division,
Los Alamos National Laboratory, Los Alamos, NM 87545, USA}

\author{Timothy C. Germann}
\affiliation{Theoretical Division,
Los Alamos National Laboratory, Los Alamos, NM 87545, USA}

\author{Alejandro Strachan}
\affiliation{School of Materials Engineering and Birck Nanotechnology Center,
Purdue University, West Lafayette, IN 47907, USA}

\date{\today}

\begin{abstract}

Mesoscale phenomena---involving a level of description between the finest  
atomistic scale and the macroscopic continuum---can be studied by a  
variation on the usual atomistic-level molecular dynamics (MD) simulation technique. In  
mesodynamics, the mass points, rather than being atoms, are mesoscopic  
in size, for instance representing the centers of mass of polycrystalline grains or  
molecules. In order to reproduce many of the overall features of fully  
atomistic MD, which is inherently more expensive, the equations of  
motion in mesodynamics must be derivable from an interaction potential  
that is faithful to the compressive equation of state, as well as to  
tensile de-cohesion that occurs along the boundaries of the mesoscale  
units. Moreover, mesodynamics differs from Newton's equations of  
motion in that dissipation---the exchange of energy between mesoparticles and 
their internal  degrees of freedom (DoFs)---must be described, and so should 
the transfer of energy between the internal modes of neighboring mesoparticles. 
We present a formulation where energy transfer between the internal modes of 
a mesoparticle and its external  center-of-mass DoFs occurs in the 
phase space of mesoparticle  coordinates, rather than momenta, resulting in 
a Galilean invariant formulation that conserves total linear momentum 
and energy (including the energy internal to the  mesoparticles). We 
show that this approach can be used to describe, in addition to mesoscale
problems, conduction electrons in atomic-level simulations of metals, and
we demonstrate applications of mesodynamics to shockwave propagation and thermal transport. 

\end{abstract}


\maketitle

\section{Introduction}

The foundations for simulating the complex motions of atoms on the  
computer by molecular dynamics (MD) are well established and go back
more than half a century.\cite{alder,vineyard} Once the interatomic interactions  
are specified, initial conditions and boundary conditions are imposed  
 the classical equations of motion---two first-order ordinary  
differential equations (Hamilton's) for coordinates and velocities, or  
one second-order ordinary differential equation (Newton's) for  
coordinates---are solved (most commonly by finite central differences).  
In MD, flows of momentum and energy in the neighborhood of any given  
atom are a natural outcome of the solution to the equations of motion.  
External driving forces can be imposed to mimic the non-equilibrium  
conditions in real-world laboratory experiments, in what has become  
known as non-equilibrium molecular dynamics (NEMD). The process of  
thermal equilibration can be viewed as either an equilibrium process  
(fluctuation dissipation) or as a consequence of non-equilibrium  
driving, and can be modeled by a heat bath with an infinite number of  
degrees of freedom (DoFs), compared to the finite number in the  
fundamental computational cell, at some fixed temperature $T_0$.

Whether equilibrium MD or NEMD, one measure of the fundamental scale of  
distance is the average nearest-neighbor distance between atoms $r_1$  
(at zero pressure and temperature, this is denoted as the bond distance  
$r_0$), and the fundamental scale of time is $r_1/c$, where $c$ is the  
speed of sound in the material being simulated (usually a dense fluid  
or solid, since dilute gases, with their rare collisions, are better modeled
via techniques such as direct simulation Monte Carlo\cite{DSMC} than
by MD or NEMD simulations). The largest distance  
scale in such simulations is the computational box size $L$, and a  
characteristic simulation time is the sound-traversal time $L/c$.  
Large-scale MD simulations are currently capable of $L/r_1 \sim 1000$
(about a billion particles in system size), for computational times that
correspond to nanoseconds of physical time.\cite{GK08}

To put the atomistic simulations into perspective, the tiny scale of  
interatomic distances dictates the maximum system size of a chunk of  
material that can be studied: at normal solid densities (zero  
temperature and pressure) in metals, $r_1 \sim 0.3$nm, so that at most,  
$L \sim 0.3\mu$m. Atomistic MD is often referred to as ``microscopic'' 
(perhaps it would be better to call it ``nanoscopic''). Typical grain  
sizes (tens to hundreds of $\mu$m) in polycrystalline metals---i.e.,
the characteristic spacings of defects---are at least 100 times larger
than the largest MD simulation currently  possible. This ``mesoscopic'' 
regime is the intermediate domain between microscopic (atomistic) and 
``macroscopic'' (continuum), where the  typical size of a finite spatial mesh 
in continuum engineering  calculations is often another factor of 100 larger still.

At a somewhat more modest scale up from atoms, one can view molecules  
as the fundamental units, rather than the atoms that make them up. In  
that case, ``mesoscopic'' can be applied in the sense that one hopes to  
achieve an approximate dynamical description that is more easily  
affordable than the fully atomistic simulation.

Finally, at the atomistic level, the non-equilibrium relaxation of  
temperature between ionic and electronic DoFs in metallic systems can  
also be viewed as ``mesoscopic.'' The thermal and transport roles of 
conduction-band electrons are completely ignored in standard MD,
but we will show that a mesodynamical formulation can be used to
incorporate their electronic effects into atomistic simulations. Whereas 
mesodynamics is meant to be  a shortcut to fully atomistic MD, whether 
in describing polycrystalline grains or molecules, in the case of electron 
thermal conduction, the result will be to add computational baggage 
(and therefore, complexity, time, and cost) to standard MD, in the hopes of 
adding important new physics, namely, a more accurate description of heat 
conduction.

Thus, the general goal of mesodynamics is to include, at least partially, 
the effects (e.g., thermal and transport) of the implicit DoFs on the dynamics 
of the explicit ones in a computationally tractable way, so that the fundamental 
mass points can represent much larger entities. Past efforts, for example, have 
focused on a minimalist formulation of the  effective interactions of the mesoparticles, 
treating the energy exchange between external and internal DoFs as relative-velocity  
damping appropriate to zero temperature \cite{holian2003}. Other mesoscopic work 
has included two-temperature MD descriptions of ions in metals coupled to a continuum 
mesh for the electronic DoFs \cite{landman} and united-atom models of molecular groups 
in polymers \cite{UnitedAtom}. The dynamics with implicit degrees of freedom (DID) model
was proposed to capture the energy exchange between the DoFs described explicitly and 
their internal modes.\cite{prl} DID was applied to shock propagation,\cite{prl} dynamical failure 
\cite{lynch09} and thermal transport \cite{zhou09} 
in molecular systems and recently generalized to describe chemical reactions \cite{antillon14}. 
Dissipative Particle Dynamics (DPD)-motivated description of mesoparticles with internal degrees of 
freedom that has been applied to both nonreactive \cite{Stoltz06} and reactive \cite{J-B07} shock waves.

In this work, we will discuss the general theory of DID mesodynamics (next Section) and results (in the 
following two Sections), followed by conclusions and prospects for the future.

\section{General Theory of Mesodynamics}

The interactions between mesoparticles can most simply be represented  
by spherically symmetric, pairwise-additive potentials; shape and  
internal symmetries could, in principle, be incorporated by adding  
non-spherical, and many-body interactions. The minimalist  
approach requires two constraints on the mesopotential  
\cite{holian2003}: (1) The compressive non-linear elastic equation of  
state should be the same as in the fully atomistic case---in other  
words, regardless of scale, bulk sound waves should travel at the same  
speed, whether at the nanoscale or at the mesoscale. (2) The cohesion  
between mesoparticles, whether polycrystalline grains or molecules,  
whose average spacing is denoted by $r_0$ (note that this is now a {\it  
mesoscopic} ``bond distance,'' rather than nanoscopic or atomistic, as  
in the Introduction), should be reduced on a per-atom basis by the  
surface-to-volume ratio $1/r_0$ (local bonds are broken  
at the surface, rather than throughout the interior). By elastic  
continuity, the compressive and tensile portions of the mesopotential  
are then coupled, with the result that the tensile stress to failure  
between mesoparticles scales with mesoparticle spacing (size) as  
$1/\sqrt{r_0}$, in accordance with the well-known Hall-Petch (or  
Griffith) relation of materials science \cite{holian2003}.

In this paper, we focus our attention upon the exchange of energy  
between the external environment of the mesoparticle, interacting  
through the mesopotential and its internal or implicit DoFs. We also 
imagine that there are cases  (such as electronic heat conduction in metals) 
where energy can flow  between neighboring mesoparticles's internal DoFs, 
so we shall include that possibility in the formalism.

The local external (or mesoparticle) velocity $\langle {\bf u} \rangle_i$ 
in the neighborhood of mesoparticle $i$ (whose own velocity is ${\bf u}_i$) 
can be obtained by averaging over its  neighbors, \cite{prl} 
using a localized, short-range weighting function $w$, reminiscent of  
smooth particle applied mechanics (or hydrodynamics)  
\cite{hoover,monaghan} or the electronic density in the embedded-atom 
method \cite{baskes}:

\begin{equation}\label{extvel}
\langle {\bf u} \rangle_i = \frac{\sum_j w(r_{ij})m_j{\bf u}_j}{\sum_j  
w(r_{ij})m_j}\ ,
\end{equation}

\noindent where $m_i$ is the mass of the mesoparticle $i$ and $r_{ij}$ is  
the distance to its neighbor $j$. The external temperature $T^{ext}_i$  
is analogously defined in $d$ spatial dimensions as

\begin{equation}\label{exttemp}
dkT^{ext}_i = \frac{\sum_j w(r_{ij})m_j|{\bf u}_j-\langle {\bf u}  
\rangle_i|^2}{\sum_j w(r_{ij})}\ ,
\end{equation}

\noindent where $k$ is Boltzmann's constant.
A typical functional form for the weighting function might be something  
like the following quartic form: 
\begin{eqnarray}\label{w}
w(r) = 
\begin{cases}
(1 - r^2/r_{max}^2)^2\ & \text{if $r < r_{max}$,} \\
0 & \text{if $r \ge r_{max}$}\ .
\end{cases}
\end{eqnarray}

\noindent Here $r_{max}$ is the range of the weighting function (typically
$r_{max}$ is similar to the range of the mesopotential). 
If $w(r_{ij}) \equiv 1$ everywhere over the entire  
computational box, the standard expressions for the system 
center-of-mass (c.m.)  
velocity and temperature, familiar to MD, are recovered. If, on the  
other hand, $w(r_{ij}) = \delta_{ij}$, then $\langle {\bf u} \rangle_i  
= {\bf u}_i$, and $T^{ext}_i \equiv 0$ (this condition prevails in the  
ballistic limit when a particle is in free-flight). Obviously, for  
purposes of mesodynamics, neither extremely long-range nor extremely  
short-range weighting functions are useful, and the range of the
mesopotential is also only a few (at most) mesoparticle spacings.

The total energy, kinetic plus potential, of $N$ mesoparticles in volume $V$,
including the energy of the internal DoFs, is

\begin{equation}  
E = \frac{1}{2}\sum_{i=1}^N m_i |{\bf u}_i|^2 + \Phi(\{{\bf r}\}) + \sum_{i=1}^N E^{int}_i\ ,
\end{equation}

\noindent and the total external force on mesoparticle $i$ is

\begin{equation}
{\bf F}_i = -\frac{\partial \Phi}{\partial {\bf r}_i}\ .
\end{equation}

We model the energy internal to mesoparticle $i$ as a general function of the internal temperature 
$T^{int}_i$: $E^{int}_i(T^{int}_i)$. The internal energy function depends on the nature and number of
internal degrees of freedom. For example, for a set of $N$ atoms within the classical harmonic 
approximation the energy would be 3$Nk$. The heat capacity for mesoparticle $i$ is defined as:

\begin{equation}\label{specificheat}
C_i(T^{int}_i) = \frac{dE^{int}_i} {dT^{int}_i}.
\end{equation}

Thus, the rate of change of the internal temperature and energy are related by:

\begin{equation}\label{tintdot}
\dot{T}^{int}_i = \frac{\dot{E}^{int}_i}{C_i}\ .
\end{equation}

The equations of motion that govern the dynamics of the mesoparticles (coordinate  
${\bf r}_i$ and velocity ${\bf u}_i$) can be written in Hamilton's form, 
with the possible addition of terms to account for the role of the internal degrees of 
freedom---a terminal  velocity ${\bf v}_i$ and a viscous deceleration ${\bf g}_i$:

\begin{eqnarray}\label{Hamilton}
{\bf \dot{r}}_i &=& {\bf u}_i + {\bf v}_i\ ,\nonumber\\
{\bf \dot{u}}_i &=& \frac{{\bf F}_i}{m_i} - {\bf g}_i\ .
\end{eqnarray}

\noindent A reasonable candidate for the dissipative terminal velocity  
is motivated by overdamped motion, where

\begin{eqnarray}\label{terminal}
{\bf \ddot{r}}_i &=& 0 = \frac{{\bf F}_i}{m_i} - \gamma {\bf  
v}_i\nonumber\\
\Rightarrow {\bf v}_i &=& \frac{{\bf F}_i}{\gamma m_i} = \chi_i {\bf  
F}_i\ .
\end{eqnarray}

\noindent A reasonable candidate for the dissipative viscous  
deceleration is Firsov relative-velocity damping \cite{firsov}:

\begin{equation}\label{damping}
{\bf g}_i = \gamma_i \left( {\bf u}_i - \langle {\bf u} \rangle_i  
\right) \ .
\end{equation}

We require that the total energy, of the external and internal DoFs of all the 
mesoparticles in the system, be conserved; i.e. the change in energy in the
mesoparticles must be exactly cancelled by that of the internal DoFs. Thus, from 
the equations of motion (Eq.\ref{Hamilton}), we see that

\begin{eqnarray}\label{edot}
\dot{E} &=& 0 = \sum_{i=1}^N \left(m_i {\bf \dot{u}}_i \cdot {\bf u}_i + \frac{\partial \Phi}{\partial {\bf r}_i} \cdot {\bf \dot{r}}_i + \dot{E}^{int}_i\right)\nonumber\\
&=& \sum_i \left(({\bf F}_i - m_i{\bf g}_i) \cdot {\bf u}_i - {\bf F}_i \cdot ({\bf u}_i + {\bf v}_i) + \dot{E}^{int}_i\right)\nonumber\\
&=& \sum_i \left(\dot{E}^{int}_i - m_i{\bf g}_i \cdot {\bf u}_i - {\bf F}_i \cdot {\bf v}_i\right)\ .
\end{eqnarray}

\noindent If we assume that the mesoparticles are $N$ independent anharmonic oscillators (Einstein model), 
which exchange energy with their neighbors only through internal DoFs, then the rate of change of the internal
energy of mesoparticle $i$ is simply given by:

\begin{equation}\label{intenergydot}
\dot{E}^{int}_i = m_i{\bf g}_i \cdot {\bf u}_i + {\bf F}_i \cdot {\bf v}_i + \nu_0 \xi_i\ ,
\end{equation}

\noindent where $\xi_i = \sum_{j \neq i} \xi_{ij}$ is the sum of internal energy transfers to 
mesoparticle $i$ from its neighbors $j$; $\nu_0$ is an arbitrary coupling rate (if $\nu_0 = 0$, 
then there are no exchanges among the internal DoFs of neighboring pairs of mesoparticles). 
Detailed balance requires that $\xi_{ji} = - \xi_{ij}$, so that summing up Eq.\ref{intenergydot} 
over all mesoparticles satisfies global energy conservation, Eq.\ref{edot}:

\begin{eqnarray}\label{sumxi}
\sum_i \xi_i &=& \sum_i \sum_{j \neq i} \xi_{ij} = \sum_j \sum_{i \neq j} \xi_{ji}\nonumber\\
&=& -\sum_i \sum_{j \neq i} \xi_{ij} = -\sum_i \xi_i = 0\ .
\end{eqnarray}

\noindent (The first step in Eq.\ref{sumxi} is exchange of dummy indices 
$ij$; the second is imposing detailed balance.)

\subsection{Ballistic and Galilean constraints}

At this stage in the development of the equations of motion for the  
mesoparticle, including both the external and internal DoFs, we  
consider two constraints that must be satisfied. First is the ballistic  
limit: if a mesoparticle should break loose from the bulk, it  
ought to travel in a straight line, since no other particles exert a  
force on it: ${\bf F}_i = 0 \Rightarrow {\bf \dot{u}}_i = 0 \Rightarrow  
{\bf \dot{r}}_i = {\bf u}_i = \text{const.}$ Thus, the dissipative  
velocity and deceleration should have no effect on the straight-line  
motion in the ballistic limit: ${\bf v}_i = 0 = {\bf g}_i$. Both the  
overdamped terminal velocity (Eq.\ref{terminal}) and the Firsov damping  
deceleration (Eq.\ref{damping}) satisfy the ballistic limit (see  
Eqs.\ref{extvel} and \ref{exttemp} and comments following Eq.\ref{w}).

The second constraint, Galilean invariance, is more stringent: the  
addition of a constant velocity (${\bf u}_p = \text{const.}$) to every  
mesoparticle (${\bf u}_i^{\prime} = {\bf u}_i + {\bf u}_p$) must alter  
neither the velocity-update equation of motion (${\bf  
\dot{u}}_i^{\prime} = {\bf \dot{u}}_i$) nor the rate of change of the  
energy of the mesoparticle's internal DoFs ($\dot{E}^{int\ \prime}_i =  
\dot{E}^{int}_i$). We require that ${\bf v}_i$ and ${\bf g}_i$, which  
augment the Hamiltonian equations of motion (Eqs.\ref{Hamilton}), each  
satisfy Galilean invariance. For example, the overdamped terminal  
velocity (Eq.\ref{terminal}) and the Firsov damping deceleration  
(Eq.\ref{damping}) are both Galilean invariant, in and of themselves.  
Obviously, the coordinate-update equation of motion is altered by the  
addition of the frame-of-reference velocity, ${\bf \dot{r}}_i^{\prime}  
= {\bf u}_i^{\prime} = {\bf \dot{r}}_i + {\bf u}_p$ (hence, ${\bf  
r}_i^{\prime} = {\bf r}_i + {\bf u}_p t$), but any {\it relative}  
coordinate is unaffected, ${\bf r}_{ij}^{\prime} = {\bf r}_{ij}$, so  
that forces are Galilean invariant: ${\bf F}_i^{\prime} = {\bf F}_i$.  
Since, by definition,  the internal temperatures do not depend on  
velocities, the exchanges among the internal DoFs of neighbors and the  
mesoparticle itself are automatically Galilean invariant.

Hence, the rate of change of the internal energy of the mesoparticle  
provides the critical constraint on the kind of dissipative terms that  
can be added to the mesodynamics equations of motion:

\begin{eqnarray}\label{Galileo}
\dot{E}^{int\ \prime}_i &=& m_i{\bf g}_i^{\prime} \cdot {\bf  
u}_i^{\prime} + {\bf F}_i^{\prime} \cdot {\bf v}_i^{\prime} +
\dot{\xi}_i^{\prime}\nonumber\\
&=& m_i{\bf g}_i \cdot ({\bf u}_i + {\bf u}_p) + {\bf F}_i \cdot {\bf  
v}_i + \dot{\xi}_i \nonumber\\
&=& \dot{E}^{int}_i + m_i{\bf g}_i \cdot {\bf u}_p\nonumber\\
&\Rightarrow& {\bf g}_i \equiv 0\ .
\end{eqnarray}

\noindent In other words, within the proposed framework, in order to satisfy Galilean invariance, we  
are only free to add a dissipative terminal velocity to the  
coordinate update; Firsov damping, even though the deceleration itself  
is Galilean invariant, cannot be combined with internal-external energy  
exchange at finite temperature in mesodynamics. Other approaches 
couple the damping to the momentum update equations and construct damping 
term in the relative velocity frame, in order to make the model Galilean invariant; 
see for example Maillet {\it et al.}\cite{J-B07}.

Thus, we propose the following equations of motion for thermo-mechanical mesodynamics 
(so-called Dynamics with Implicit DoFs - DID \cite{prl,lynch09}):

\begin{eqnarray}\label{DID}
{\bf \dot{r}}_i &=& {\bf u}_i + \chi_i{\bf F}_i\ ,\nonumber\\
{\bf \dot{u}}_i &=& \frac{{\bf F}_i}{m_i}\ ,\nonumber\\
\dot{E}^{int}_i &=& \chi_i|{\bf F}_i|^2 + \dot{\xi}_i\ .
\end{eqnarray}

\subsection{Energy exchange using direct and integral feedback}

The feedback term $\chi_i$ that appears in Eqs. \ref{DID} can either  
be direct, as in the Berendsen style of thermostatting  
\cite{Berendsen}, or integral, as in the Nos\'e-Hoover style. \cite{NH}
Integral feedback is time-{\it reversible}, and requires an additional  
heat-flow variable for each local, finite thermostat, whose equation of  
motion guarantees, in the long-time average sense, that the two  
temperatures being thermostatted will equilibrate with each other, even  
in the non-equilibrium steady state. Under time-reversible equations of  
motion, if a system is first integrated forward in time from zero to  
$t$, followed by reversal of time and velocities (including the  
auxiliary flow variables), such that $t \rightarrow -t$ and ${\bf u}_i  
\rightarrow -{\bf u}_i$, and then the system is integrated forward for  
a time $t$, it will, in principle (within a Lyapunov time), return to  
its initial condition at $t=0$. Newton's equations of motion are  
time-reversible, as are these for integral-feedback mesodynamics.  
Direct feedback, on the other hand, is time-{\it irreversible} and
it does not require any additional heat-flow variable.
But direct feedback does not  
absolutely guarantee that the two temperatures come to exact  
equilibrium with each other, particularly under external driving.
\cite{holian95}

We can write the feedback term formally as

\begin{equation}\label{chi}
\chi_i = \frac{\nu \zeta_i}{m_i \omega_E^2}\ ,
\end{equation}

\noindent where $\nu$ is the coupling rate of the internal-external  
energy-exchange thermostat to the mesoparticle, $\zeta_i$ is the  
dimensionless heat-flow variable (positive means that heat flows from  
the mesoparticle's external motion into the internal DoFs), and  
$\omega_E = \overline{\omega^2}$ is the Einstein frequency of the  
mesoparticle's external DoFs, defined as the long-time average of the  
trace of the dynamical matrix:

\begin{equation}\label{om2}
\omega^2 = \frac{1}{dN}\ \sum_{i=1}^N\  
\frac{1}{m_i}\frac{\partial}{\partial {\bf r}_i} \cdot \frac{\partial  
\Phi}{\partial {\bf r}_i}\ .
\end{equation}

\noindent Notice that we have chosen this particular parameterization  
of the feedback with malice aforethought: if $\nu = 0$ then Newton's  
(Hamilton's) equations of motion are recovered; also, the units are  
such that $\chi_i {\bf F}_i$ is a velocity. (Of course, in practical  
applications, the value of the Einstein frequency can be guessed at,  
since there is some arbitrariness inherent in the choice of the  
coupling rate $\nu$.)

The integral feedback equation of motion for the heat-flow variable is  
given by

\begin{equation}\label{zetadot}
\dot{\zeta}_i = \nu \left(\frac{T^{ext}_i-T^{int}_i}{T_0}\right)\ ,
\end{equation}

\noindent where $T_0$ is an arbitrary temperature (for example, a guess  
at the final equilibrium value). The long-time average of this equation  
of motion is zero; hence, at long times, $\overline{T^{ext}_i} =  
\overline{T^{int}_i}$ (either at equilibrium or at the non-equilibrium  
steady state).

In order to obtain the direct-feedback form of our thermo-mechanical  
mesodynamics equations of motion, we simply substitute for $\nu  
\zeta_i$ the expression for $\dot{\zeta}_i$ (Eq.\ref{zetadot}) into the  
expression for $\chi_i$ (Eq.\ref{chi}); that is,

\begin{equation}\label{chit}
\tilde{\chi}_i = \frac{\nu}{m_i \omega_E^2}
\left(\frac{T^{ext}_i-T^{int}_i}{T_0}\right)\ .
\end{equation}

\subsection{Exchange of internal temperature among neighbors}

We now complete the thermo-mechanical description of the mesodynamics  
equations of motion (Eqs. \ref{DID}) by deriving the equation of  
motion for the transfers of internal energy between neighbors of  
mesoparticle $i$ (see Eq.\ref{intenergydot} and comments thereafter) as the time-reversible version of  
Fourier's Law ($\kappa$ is the thermal conductivity):

\begin{equation}\label{Fourier}
\dot{\xi_i} = \kappa \ \nabla^2 T^{int}_i\ .
\end{equation}

\noindent The exchange of internal energy among pairs of neighboring  
mesoparticles is particularly important when $\kappa$ is large, as in  
the case of electrons in metals, where the mesoparticles are the ions  
themselves, and the ion's share of conduction electrons in its vicinity  
serve as its internal DoFs.

We can express the Laplacian of the internal temperature in  
Eq.\ref{Fourier} in the following form, which is consistent with the  
finite-difference expression for a lattice. In our implementation, however, 
the mesoparticles themselves form the Lagrangian ``mesh'' (similar to SPAM  
\cite{hoover}) over which the Laplacian is evaluated:

\begin{equation}\label{Laplace}
\nabla^2 T^{int}_i = \alpha \sum_{j \neq i} w(r_{ij}) \ \frac{T^{int}_j  
- T^{int}_i}{r^2_{ij}}\ .
\end{equation}

\noindent The normalization factor $\alpha$ is determined from an  
arbitrary analytic quadratic temperature profile in a reference lattice  
(in $d$ dimensions, with $N_n$ neighbors at distance $r_n$, for up to  
$n_{max}$ shells of neighbors, such that $r_{n_{max}} \leq r_{max}$):

\begin{equation}\label{normal}
\alpha = 2d \ \Big{/} \ \sum_{n=1}^{n_{max}}\ N_n w(r_n)\ .
\end{equation}

Thus, the rate of internal energy transfer between pairs of  
mesoparticles can be expressed as

\begin{equation}\label{xidot}
\dot{\xi}_{ij} = \kappa \alpha w(r_{ij}) \ \frac{T^{int}_j -  
T^{int}_i}{r^2_{ij}}\ ,
\end{equation}

\noindent which exhibits detailed balance in differential form  
(it is antisymmetric under exchange of the pair indices ($\dot{\xi}_{ji} =  
- \dot{\xi}_{ij}$), so that whatever is lost from one particle to  
another is gained by the other. This guarantees global energy  
conservation for $N$ mesoparticles in volume $V$.

The thermo-mechanical mesodynamics equations of motion for integral  
feedback (Nos\'e-Hoover-style) can then be summarized as follows:

\begin{eqnarray}\label{SHEOM_integral}
{\bf \dot{r}}_i &=& {\bf u}_i + \frac{\nu \zeta_i}{m_i \omega_E^2}{\bf  
F}_i\ ,\nonumber\\
{\bf \dot{u}}_i &=& \frac{{\bf F}_i}{m_i}\ ,\nonumber\\
\dot{E}^{int}_i &=& C^{int}_i T^{int}_i \nonumber \\
                &=& \frac{\nu \zeta_i}{m_i \omega_E^2}|{\bf F}_i|^2 +  
\nu_0 \xi_i\ ,\nonumber\\
\dot{\zeta}_i &=& \nu \left(\frac{T^{ext}_i-T^{int}_i}{T_0}\right)\  
,\nonumber\\
\dot{\xi_i} &=& \kappa \ \nabla^2 T^{int}_i\ .
\end{eqnarray}

\noindent For direct feedback (Berendsen-style), the equations of  
motion can be summarized as follows:

\begin{eqnarray}\label{DID_direct}
{\bf \dot{r}}_i &=& {\bf u}_i + \frac{\nu}{m_i \omega_E^2}
\left(\frac{T^{ext}_i-T^{int}_i}{T_0}\right){\bf F}_i\ ,\nonumber\\
{\bf \dot{u}}_i &=& \frac{{\bf F}_i}{m_i}\ ,\nonumber\\
\dot{E}^{int}_i &=& C^{int}_i T^{int}_i \nonumber \\
                &=& \frac{\nu}{m_i \omega_E^2}
\left(\frac{T^{ext}_i-T^{int}_i}{T_0}\right)|{\bf F}_i|^2 + \kappa \  
\nabla^2 T^{int}_i\ .\nonumber\\
\end{eqnarray}

\noindent (See Appendix B for the finite central-difference realization  
of these mesodynamics equations of motion.) In the remainder of the  
paper, we will apply mesodynamics to a variety of themo-mechanical  
problems.

\section{Shock loading of a mesoscale model of unreactive HMX}

As a first DID example we study shock propagation in a model molecular crystal. 
We focus on the exchange of energy between the mesoparticles and their internal 
DoFs as well as on the role of quantum effects on the specific heat
of the internal DoFs in the prediction of the temperature of the shocked 
material.

Mesoparticle approaches are widely used to describe molecular materials (e.g. polymers,
biomolecules, molecular crystals). However, these approaches often treat the thermal role of the 
implicit degrees of freedom very crudely or disregard it altogether. We now apply DID
to simulate the propagation of a shockwave in a model molecular crystal
based on the properties of the high-energy density material  HMX (cyclic [CH$_2$-N(NO$_2$)]$_4$) focusing on the role of the thermal properties of the internal DoFs.

The propagation of shockwaves in molecular crystals is very challenging to 
mesodynamics, since large amounts of energy are exchanged in very short
time-scales.\cite{prl,lynch09,J-B07} The translational energy in the shockwave initially excites  
long-wavelength, low-energy intermolecular DoFs (the ones described  
explicitly at the mesoscale), resulting in short-lived overheating
of these (few) modes. \cite{prl,jaramillo07} 
Part of this energy then ``cascades'' to higher-energy, higher-frequency intramolecular 
DoFs, which are only  implicitly treated in mesoscopic descriptions; this process 
occurs over a short time-scale that depends on the details of the molecular vibrational 
spectrum. This equilibration process continues until the inter- and intramolecular 
DoFs attain the same temperature. The final temperature of the shocked
material depends on the equation of state of the material and the specific
heat plays a critical role determining how much of the energy deposited
in the material by the shock ends up being kinetic energy (or temperature). 

\subsection{Model and simulation details}

In our model molecular crystal each mesoparticle represents a molecule and their 
interaction is described via a two-body Morse potential. The model system has a face centered cubic
(fcc) crystal structure and the potential was parametrized to reproduce the uniaxial compression, density, and 
cohesive energy of HMX as described in Ref. \onlinecite{lynch09}. We simulate shock propagation  
using the direct feedback flavor of mesodynamics, Eqs. \ref{DID_direct}, with no direct exchange of energy 
between the internal degrees of freedom ($\kappa$ = 0); note that the internal DoFs still interact with one 
another via the mesoparticles. The mean square frequency in Eqs. \ref{DID_direct} was obtained from the 
vibrational density of states of the model system: $\langle \omega^2 \rangle$=2.04 ps$^{-2}$.

We simulate the shock loading using non-equilibrium simulations involving high-velocity impact of the 
target material into a static piston at the desired particle (or piston) velocity ($u_p$). Galilean invariance makes
this setup equivalent to an infinitively massive piston hitting the target at velocity $u_p$.  Both
piston and target simulation cells consist of a fcc crystals oriented along x = [1 0 0], y = [0 1 0], and z = [0 0 1] 
and lattice parameters of 1.02265 nm. The piston consists of 3200 molecules (2 x 20 x 20 unit cells), and the positions
of all its molecules are fixed throughout the simulation. The target consists of 160,000 molecules (100 x 20 x 20 unit cells). 
The target and piston are initially separated by 2.7 nm along the shock direction (x). The target is thermalized at 300 K to achieve equilibrium between mesoparticles and their internal DoFs.  After this thermalization, a translational velocity 
(0.4, 1.0, and 2.0 km/s) is added to each mesoparticle over the thermal velocities. 
During these simulations, the DID coupling constant $\nu$ is taken as 0.0023 ps$^{-1}$ which leads to equilibration timescales similar 
to those in the all-atom simulations.\cite{jaramillo07} The MD timestep is 5 fs and periodic boundary conditions are applied to the 
y and z directions, and open boundary condition is applied to the x direction.  

\subsection{Shock-induced temperature increase: the classical case}

In order to compare the mesodynamical results with all-atom MD in Ref. \onlinecite{jaramillo07} 
we used the classical harmonic approximation for the specific heat: $C^{int}_i/k_B=N^{int}$, 
where $N^{int}=78$ is the number of implicit DoFs. 

\begin{figure}[htbp]
\includegraphics[width=5in]{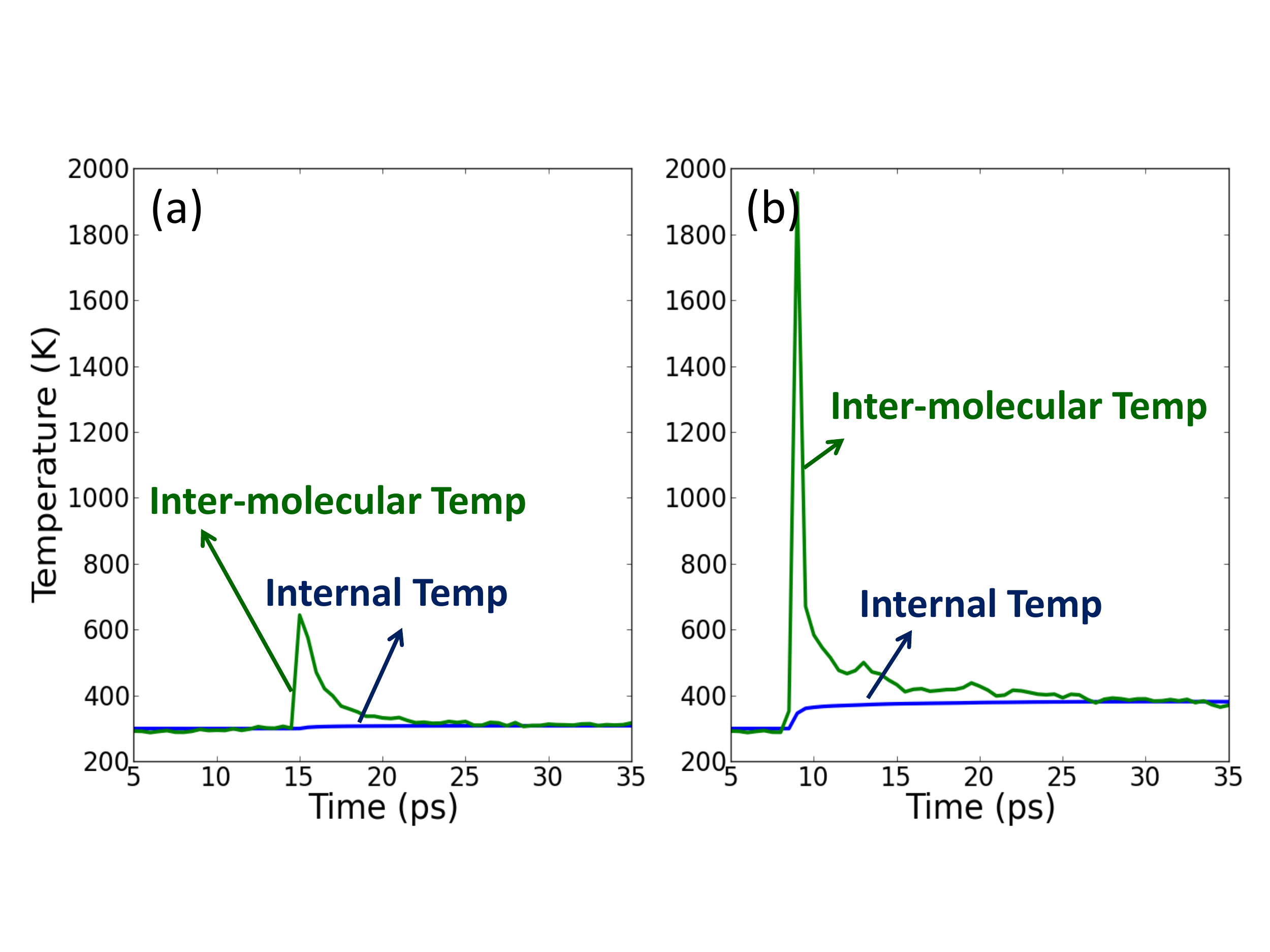}
\caption{ \label{Fig:TempCompare}
Time dependence of the local temperatures of a thin slab of HMX as a
shock passes through with piston velocity u$_p$ = (a) 0.4 km/s and (b) 1.0 km/s. We show
inter-molecular and  internal temperatures for mesodynamics simulation with coupling to internal
modes.}
\end{figure}

Figure \ref{Fig:TempCompare} shows the time evolution of the temperatures of a thin slab of material 
(one unit cell wide) for piston velocities u$_p$ of 0.4 km/s and 1.0 km/s.  
For each case we show: (i) the {\it  inter-molecular} temperature,
defined from the kinetic energy of the mesoparticles measured around the c.m. translational 
velocity of the entire slab; and (2) the {\it  internal} temperature, defined as
an average over the internal temperature of the mesoparticles in the slab. 
The DID predictions of temperature increase are in good agreement with the all-atom
simulations even when the mesoscale model is very simple. By appropriately adjusting
the DID coupling constant we also obtain good agreement regarding the equilibration
timescales between the explicit and implicit modes and on the amount of overheating
experienced by the molecular modes. \cite{jaramillo07} 

\subsection{Shock-induced temperature increase: quantum effects}

Up to this point the thermal properties of the internal degrees of
freedom have been treated classically. All-atom MD is always classical
(in equilibrium, every DoF that appears squared in the Hamiltonian 
takes 1/2k$_B$T of energy) and, in order to compare with such results,
we have used the classical harmonic approximation to obtain the specific
heat for mesodynamics. However, classical statistical mechanics significantly
overestimates the specific heat of molecular materials and we now use
DID to assess the shock-induced heating when a correct quantum description
is used.

Within the harmonic approximation, the internal modes of a molecule can be described 
by a set of N$^{int}$ independent harmonic oscillators (normal modes) characterized by 
their frequency $\omega_i$; this is usually a good approximation for the internal degrees
of freedom of a molecule in a wide temperature range and leads to analytical
expressions for the specific heat both within classical and quantum mechanics. 
Classically, the internal energy of such harmonic system is given by:
E$^{int}_{CM}$=N$^{int}$k$_B$T and, as mentioned before, the specific heat 
is C$^{int}_{CM}$=N$^{int}$k$_B$, where N$^{int}$ is the number of normal modes. 
According to QM, the energy 
of a harmonic oscillator is given by: $e_{\omega_i}(n_i) = \hbar\omega_i$ 
(1/2+n$_i$), where the quantun number n$_i$ is an integer that gives the 
excitation of mode $i$, $\omega_i$ is its frequency and $\hbar$ is Planck's
constant. Using statistical mechanics we can calculate the average QM internal 
energy of a molecule at temperature T within the harmonic approximation:
\begin{eqnarray}\label{Eq:QMenergy}
E^{int}_{QM}(T) &=& \sum_{i=1}^{N^{int}} 
               \frac{\sum_{n_i=0}^{\infty} e_{\omega_i}(n_i) \exp[-e_{\omega_i}(n_i)/kT]}
                    {\sum_{n_i=0}^{\infty} \exp[-e_{\omega_i}(n_i)/kT]} \nonumber \\
           &=& \sum_{i=1}^{N^{int}} \frac{1}{2} + \hbar\omega_i 
                 \frac{1}{\exp(\hbar\omega_i/kT) - 1},
\end{eqnarray}
the specific heat is, thus:
\begin{equation}\label{Eq:QM_Cv}
C^{int}_{QM}(T) = k_B \sum_{i=1}^{N^{int}} \frac{\left(\hbar\omega_i/kT\right)^2
                    \exp(\hbar\omega_i/kT)}{\left(1-\exp(\hbar\omega_i/kT)\right)^2}.
\end{equation}

Since each mode can only take or give energy in a quantized manner (in lumps of
$\hbar\omega_i$), high-frequency, high-energy modes are not excited at low temperatures 
and the specific heat is much smaller than the classical value. In Fig. \ref{Fig:Cv}
we show the temperature dependence of the specific heat of HMX calculated using
Eq. \ref{Eq:QM_Cv} and normal mode frequencies obtained from the experimental data
in Ref. \onlinecite{brand02}.  

\begin{figure}[htbp]
\includegraphics[width=5in]{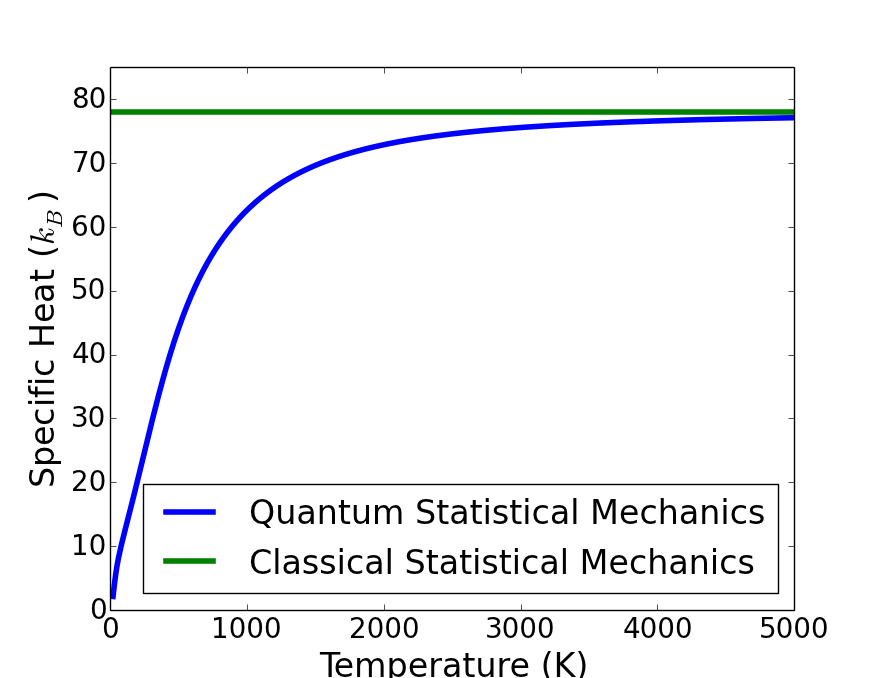}
\caption{ \label{Fig:Cv}
Quantum and classical specific heats for HMX as a function of temperature.}
\end{figure}

While the thermal properties of all-atom MD simulations is always classical,
our mesodynamic approach (Eqs. \ref{SHEOM_integral} and \ref{DID_direct}) enables the 
use of a quantum mechanically-derived, temperature dependent, specific heat 
to describe the thermal role of the internal DoFs. In Fig. \ref{Fig:Cv} we show the time 
evolution of the internal temperature of a thin HMX slab as shockwaves with 
u$_p$=0.4 and 1.0 km/s pass through it using both classical (full lines) and 
quantum (dashed lines) specific heats. As expected, the smaller value of the quantum specific heat leads 
to higher shock temperatures and this effect is more marked for weaker shocks.
For the shock with u$_p$ = 1.0 km/s the temperature increase is underestimated by
a factor of approximately two in 
a classical description (remember the mesodynamics with classical specific heat 
agrees well with all-atom MD). Interestingly, the quantum DID results are more accurate 
(within the limitations of the mesopotential and the input specific heat) than 
all-atom MD and classical mesodynamics. 

The overestimation of the specific heat of materials below their Debye temperature by classical
mechanics should be carefully acknowledged in the interpretation of MD simulations, especially 
of non-equilibrium processes like shock-induced chemical reactions \cite{strachan03} 
and thermal transport. \cite{donadio09,turney09} 

\begin{figure}[htbp]
\includegraphics[width=5in]{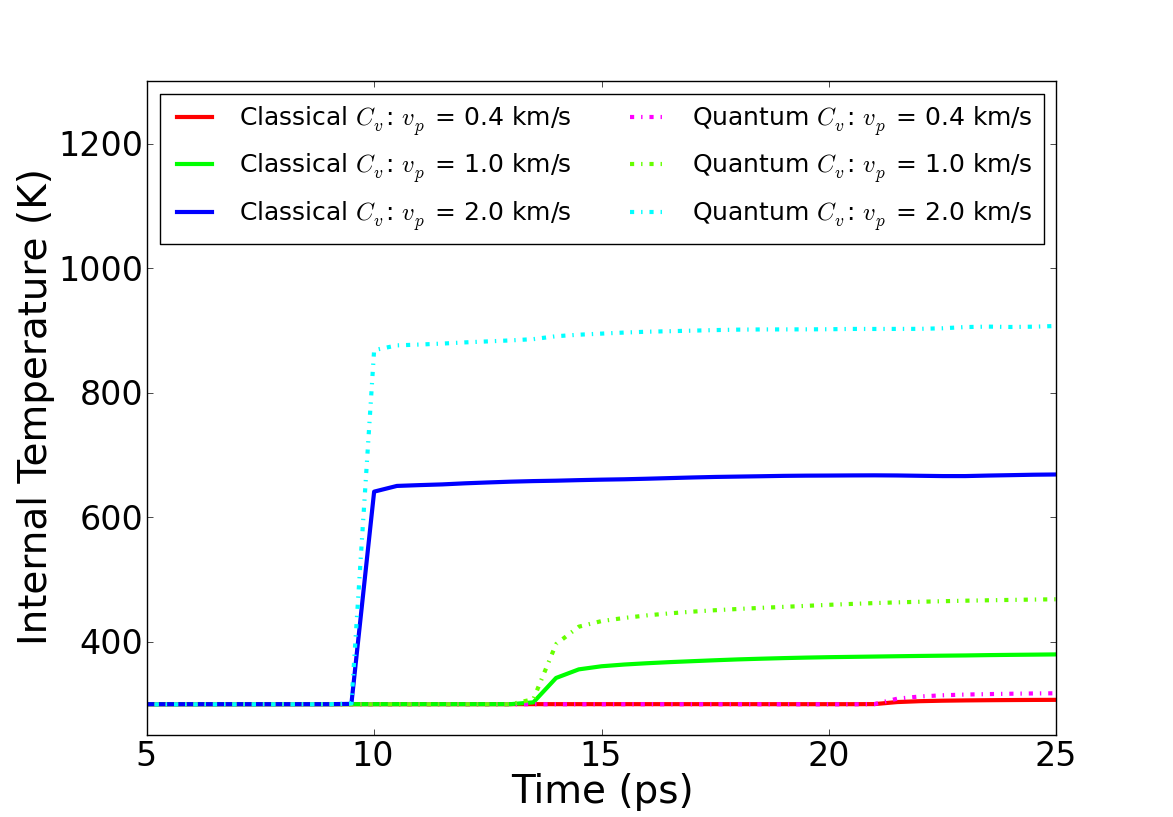}
\caption{ \label{Fig:Cv}
Time dependence of the internal temperatures of a thin HMX slab with classical and quantum specific heats and under shocks with u$_p$ = 0.4, 1.0, and 2.0 km/s.}
\end{figure}

\section{DID for electrons: thermal transport by electrons and phonons}

Understanding nanoscale heat transport in metallic systems, where conduction electrons 
play a dominant role, is important for the applications in heat-assisted magnetic recording,\cite{Kryder08}
microelectromechanical systems (MEMS),\cite{Bechtold05, Hyman99, Jensen05} high power laser,\cite{Huyeh08, Guo12} and thermoelectric devices.\cite{Harman02, Chen02} 
Challenges involved with direct experimental measurements at the nanoscale and the separation 
between the electronic and the phononic contributions make atomic simulations attractive. Standard
all-atom MD simulations provide an explicit description of phonons but ignore the transport role 
of electrons and are, therefore, incapable of capturing thermal transport in metals.

To address this limitation the MD technique has been combined with the two-temperature model 
to simulate the thermal transport in metal and metal-semiconductor systems.\cite{Rutherford07,
Duffy07,Phillips09,Wang12,Li09}
In these simulations, electronic transport is described by solving the diffusion equation over
a grid that overlaps with the atomic system. The two subsystems are coupled with local phonon
temperature obtained from the atomic velocities.
 
In this Section, we apply DID to simulate the thermal transport in metals via the two-temperature
model. Sub-section \ref{Sect:DID-Ele-Ver} shows verification tests of our implementation that also
serve the purpose of exemplifying the use of the method. Sub-section \ref{Sect:ElectronResults} 
discusses thermal transport calculations in nanoscale Al focusing on the effects of specimen 
size and electron-phonon coupling rate.   

\subsection{Simulation details}
\label{Sect:ElectronSimuDetails}

In this DID simulations particles represent Al atoms and their interaction is described by an embedded 
atom model potential \cite{Mishin02}. Conduction electrons are described as implicit DoFs. Since these 
electrons are not tied to the atoms, we include diffusive transport between them as described
by Eq. \ref{DIDELE_direct}. We repeat the equations here using {\it elec} to indicate electronic 
properties and {\it atom} for atomic ones :

\begin{eqnarray}\label{DIDELE_direct}
{\bf \dot{r}}_i &=& {\bf u}_i + \frac{\nu}{m_i \omega_E^2}
\left(\frac{T^{atom}_i-T^{elec}_i}{T_0}\right){\bf F}_i\ ,\nonumber\\
{\bf \dot{u}}_i &=& \frac{{\bf F}_i}{m_i}\ ,\nonumber\\
\dot{E}^{elec}_i &=& C^{elec}(T^{elec}_i) \dot{T}^{elec}_i \nonumber \\
                &=& \frac{\nu}{m_i \omega_E^2}
\left(\frac{T^{atom}_i-T^{elec}_i}{T_0}\right)|{\bf F}_i|^2 + \kappa^{elec} \  
\nabla^2 T^{elec}_i\ .\nonumber\\
\end{eqnarray}
In these equations, $\nu$ is the coupling rate between the electronic and atomic subsystems, 
and $\kappa^{elec} $ is the electronic thermal conductivity. Our DID approach to describe electronic 
thermal transport is similar to the MD two-temperature model\cite{Wang12,Duffy07,Rutherford07} except that the electronic 
temperatures are based on the atomic positions instead of the spatial grid and that the coupling is done via the position update equation. The benefit of using atomic positions as a 
grid for electron-phonon coupling is that the atomic structures at the interfaces and free surfaces can be captured accurately and in a straightforward manner. 

We use coupling rates $\nu$ between 0.00017 ps$^{-1}$ and 0.017 ps$^{-1}$, corresponding to electron-phonon 
coupling constants of 0.01-1$\cdot 10^{17}$ $W/m^3K$. The electronic energy per 
atom, $E^{elec}$, is set to $7.975\cdot 10^{-5}\cdot{T^2}\cdot{k_b}$\cite{Dicke67} leading to a specific heat per atom $C^{elec}$ to be $1.595\cdot 10^{-4}\cdot k_b\cdot T$. The electronic thermal conductivity, 
$\kappa ^{elec}$, is $27259.245\cdot k_b\cdot \AA^2/ps$, which is equivalent to 222 W/(mK)\cite{asm}. 
$<\omega ^{2}>$ is 39.783 ps$^{-2}$, which is obtained from the Debye temperature of Al.\cite{Bose09}

\subsection{Verification tests}
\label{Sect:DID-Ele-Ver}

The DID equations of motion conserve total energy (the sum of the electronic and atomic 
subsystems) even when the energy is exchanged between the atoms and the electrons. To verify the 
energy conservation during the equilibration process between the atomic and electronic subsystems, 
an fcc Al system  containing 4,000 atoms (5 x 5 x 40 unit cells) was used. For this system, the 
lattice constant is 0.408 nm, and the cell oriented along x = [100], y = [010], and z = [001]. 
Initial temperatures of 600K and 300K are assigned to the electronic and atomic subsystems 
and the equilibration process is followed via an isochoric-adiabatic (NVE) simulation for 20 ps.
The atomic timestep is set to 0.1 fs and 40 electronic timesteps are performed per atomic timestep. 

\begin{figure}[htbp]
\includegraphics[width=7in]{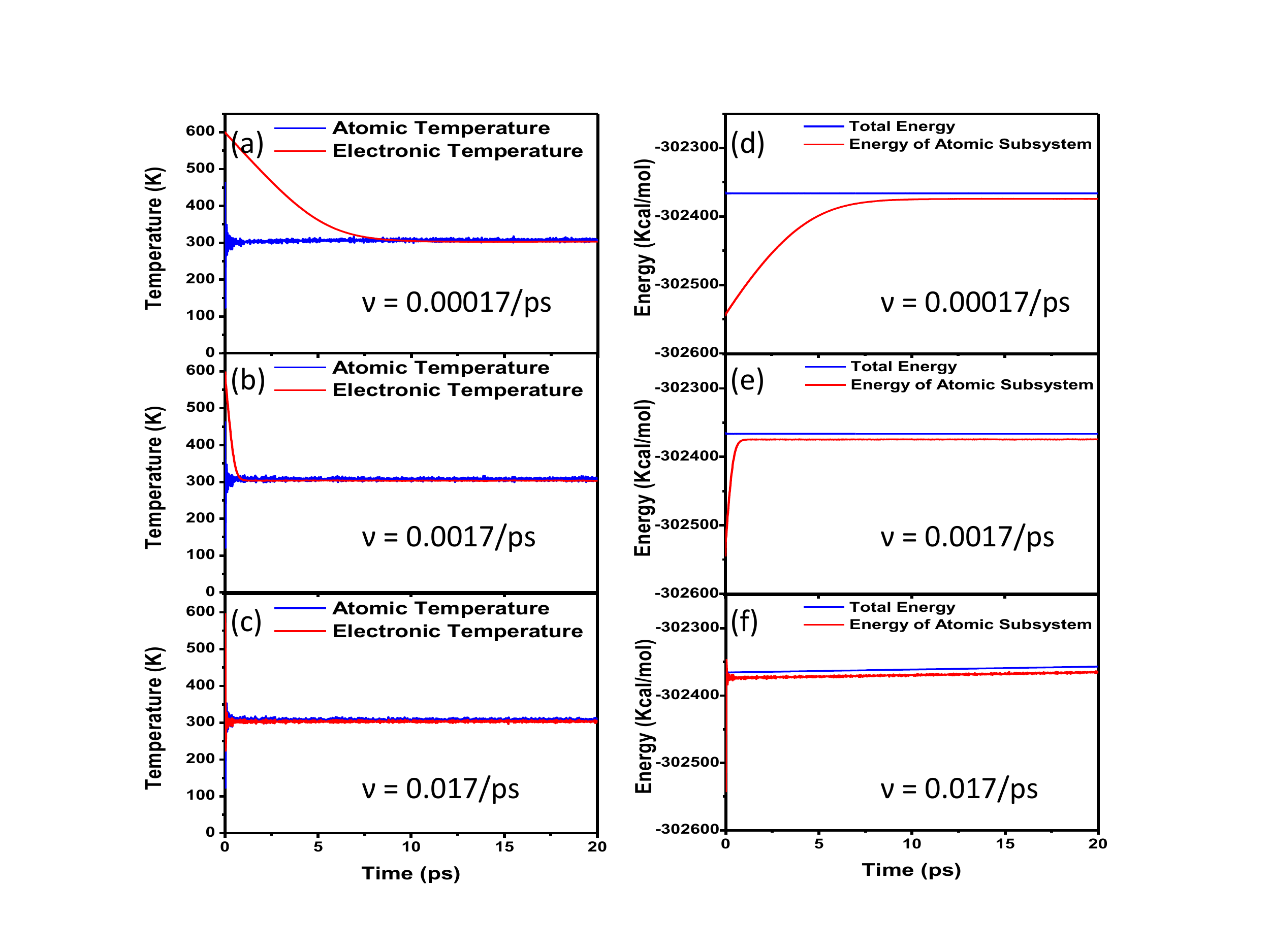}
\caption{ \label{Fig:EnergyCons}
(a,b,c) Temperature equilibration between the electronic and atomic subsystems and (d,e,f)
time evolution of the total and atomic energies for different electron-phonon
coupling constants.}
\end{figure}

The equilibration between electronic and atomic temperatures with time for various coupling constants is 
shown in Figs. \ref{Fig:EnergyCons} (a-c) and the corresponding total and subsystem energies 
in Figs. \ref{Fig:EnergyCons} (d-f). Since the specific heat of atoms is much larger than that of electrons, 
the final equilibrium temperatures are close to, but slightly larger than, the initial atomic 
temperature. The role of coupling constant in equilibration time is clear from the figure. 
From Figs. \ref{Fig:EnergyCons} (d) and (e), the total energies of the systems are conserved 
during the equilibration process for systems with $\nu$ of 0.0017 ps$^{-1}$ and 0.00017 ps$^{-1}$. However, due 
to the very large energy exchange between electrons and atoms for the system with 
$\nu$ of 0.017 ps$^{-1}$, the total energy for this system drifts during the equilibration process, 
see Fig. \ref{Fig:EnergyCons} (f). 

The time-evolution of electronic temperature is governed by the diffusion equation. To
verify our implementation we set $\nu$ to 0.00 ps$^{-1}$ to decouple the atomic and electronic
degrees of freedom and solve the time evolution of the electronic temperature given an
initial Gaussian temperature profile. The initial electronic temperature is:
\begin{eqnarray}\label{eq:Temp_Gaussian}
T_{i}^{elec}(t=0,z)=300+200\cdot exp\left ( \frac{-(203.938-z)^2}{1600} \right )
\end{eqnarray}

With this initial condition we perform a DID simulation  under isochoric-adiabatic (NVE) conditions
using a temperature-independent electronic specific heat $C^{elec} = 0.048711\cdot k_b$. The 
analytical solution of the time evolution of the temperature profile is:
\begin{eqnarray}\label{eq:Temp_sol}
T_{i}^{elec}(t,z)=300+\left ( \frac{4000}{\sqrt{\alpha }} \right )exp\left ( \frac{-(203.9-z)^{2}}{4\alpha } \right )
\end{eqnarray}
where $\alpha$ is equal to $\frac{\kappa ^{elec}}{C^{elec}}(t+t_{0})$.

\begin{figure}[htbp]
\includegraphics[width=5in]{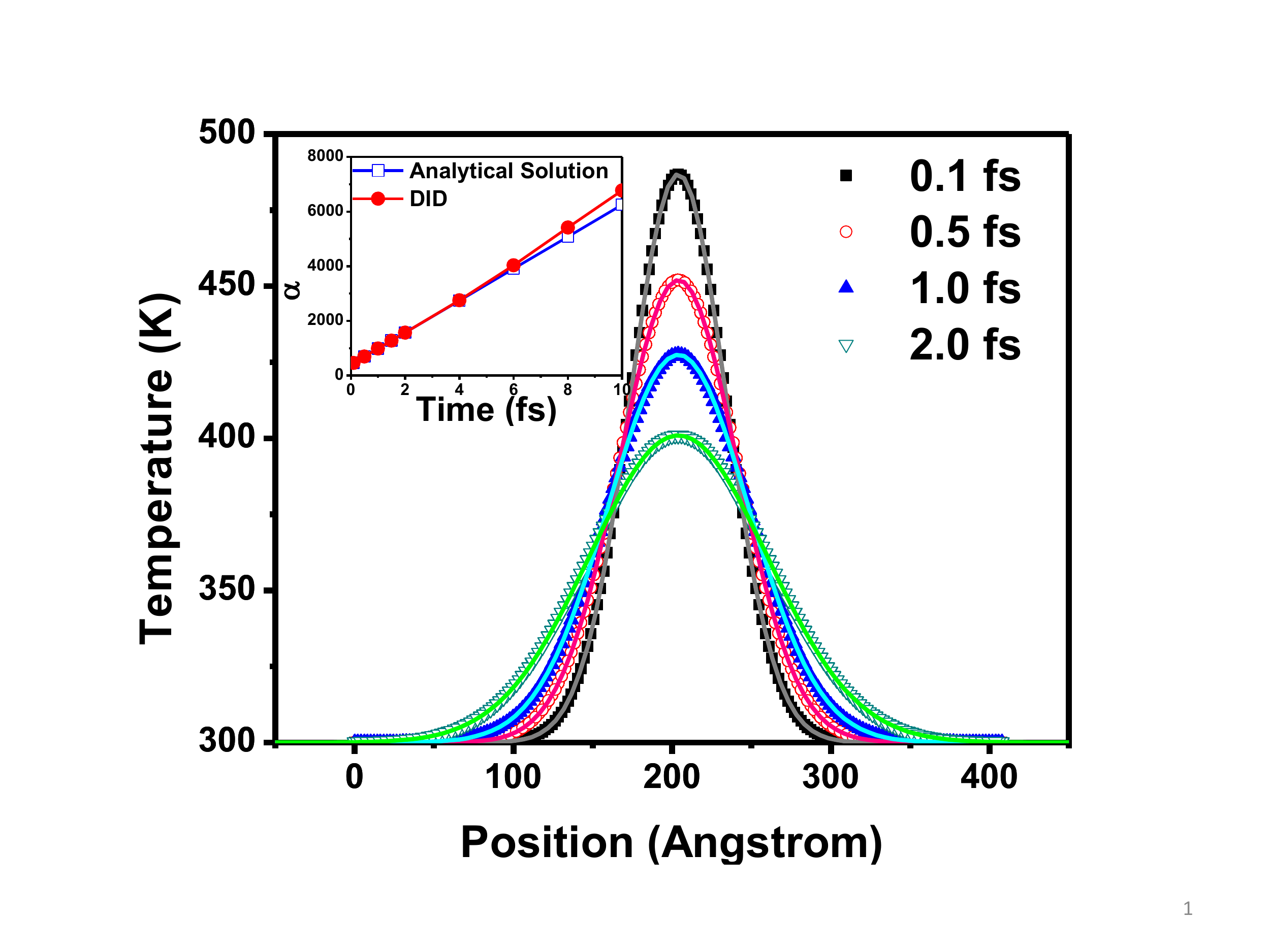}
\caption{ \label{Fig:Difussivity}
Electronic temperature profiles at different times without coupling to the atomic temperature. The points are the results from DID simulation and the solid lines are the results from analytical solution. Inset: broadening factor, $\alpha$ calculated from the analytical solution and from the electronic temperature evolution of DID simulation.}
\end{figure}

The DID temperature evolution is compared with the analytical solution in Fig. \ref{Fig:Difussivity}. The 
inset shows the width of the distribution, $\alpha$, as a function of time for both cases. In this simulation, 
the positions of moving atoms obtained from the equations of motion are used as a grid for the electronic 
temperatures, while the MD two-temperature model uses a fixed spatial grid to describe the electronic 
temperatures. The MD results agree with the analytical solutions for short simulation times when the periodic 
boundary conditions lead to negligible interactions between the temperature profiles in neighboring cells. 

\subsection{Effect of size and electron-phonon coupling constant}
\label{Sect:ElectronResults}

We now use non-equilibrium DID simulations to study the role of specimen size and electron-phonon coupling 
constant, $\nu$, on the thermal conductivity of Al. We perform simulations on 3D periodic systems with size varying from
5 x 5 x 40 unit cells (4,000 atoms) to 5 x 5 x 400 unit cells (40,000 atoms) with coupling
constants $\nu$ ranging from 0.00 ps$^{-1}$ to 0.017 ps$^{-1}$. 

We compute thermal conductivity via non-equilibrium DID simulations with a method proposed
by M$\ddot{u}$ller-Plathe method.\cite{MullerPlathe97} Heat fluxes in the range of $1-2\cdot10^{11} 
W/m^2$ are introduced by periodically swapping the atomic velocities of the coldest atom in the 
hot bin (the first bin) and the hottest atom in the cold bin (the middle bin) every 10 fs (every 100 
atomic timesteps). The total simulation time is 200 ps for systems with $\nu$ equal to 0.00 ps$^{-1}$ and 
the thermal conductivity is calculated from 100-200 ps. The total simulation time is 80 ps for systems 
with $\nu$ from 0.00017-0.017 ps$^{-1}$ and the thermal conductivity is calculated from 40-80 ps. 
In this method the thermal conductivity is computed using Fick's law as the ratio between the 
temperature gradient and the heat flux. The temperature gradient is obtained from the linear fit to the effective temperature $T^{eff}$ of the whole material excluding the hot and cold bins. $T^{eff}$ is defined by both electronic and atomic subsystems via:
\begin{eqnarray}\label{effTemp}
C^{total}(T^{eff})\cdot T^{eff}=C^{atom}\cdot T^{atom}+C^{elec}(T^{elec})\cdot T^{elec}
\end{eqnarray}
where $C^{atom}$ is the specific heat of atomic subsystem and is equal to $3\cdot k_b$ per atom, and $C^{total}$ is 
the total specific heat of the system. $T^{atom}$ and $T^{elec}$ are the temperatures of atomic and electronic subsystems, 
respectively. 

\begin{figure}[htbp]
\includegraphics[width=7in]{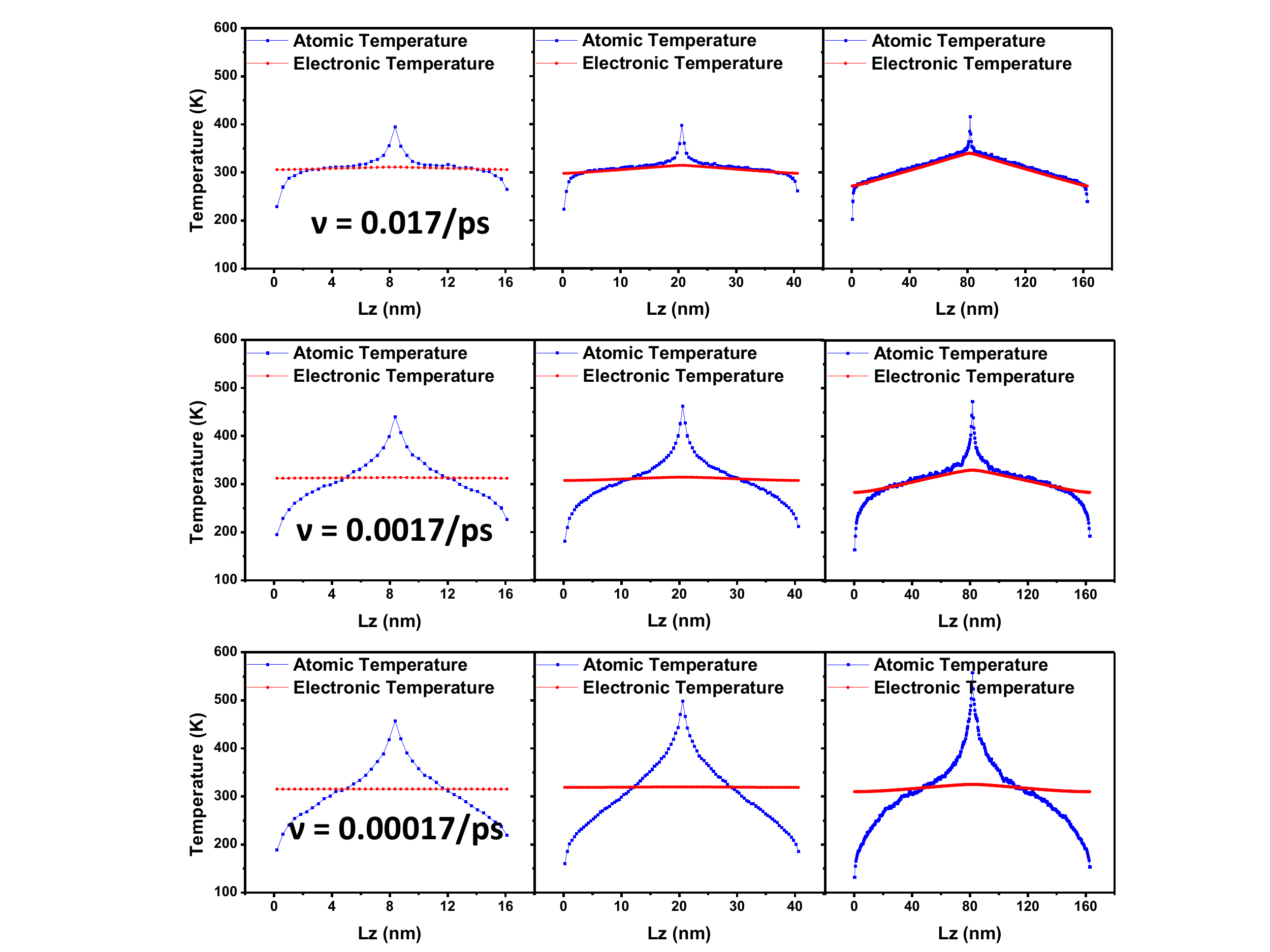}
\caption{ \label{Fig:TempProfiles}
Steady-state electronic and atomic temperature profiles during non-equilibrium MD 
simulations for various electron-phonon coupling constants and specimen sizes.}
\end{figure}

The electronic and atomic temperature profiles of Al with $\nu$ from 0.00017-0.017 ps$^{-1}$ and specimen 
lengths from 16.32-163.20 nm are shown in Fig. \ref{Fig:TempProfiles}. Due to the periodic 
kinetic energy exchange in the hot and cold bins, the temperatures of atoms and electrons 
remain out of equilibrium in these areas for all cases. For relatively long specimens
and high coupling constants the atoms and electrons come to local equilibrium away from
the energy-exchange zones, see Fig. \ref{Fig:TempProfiles}. As will be shown below the
thermal conductivity in such cases is approximately the phonon thermal conductivity
(obtained as a function of size for the $\nu$=0.00 ps$^{-1}$ simulations) plus the electron one (208 W/(mK)).
Interestingly, for smaller specimens or weaker electron-phonon coupling constants the electrons
do not reach local thermal equilibrium with the atoms and do not fully participate in heat
transport. Similar results have been shown by the theoretical calculations of 
two-temperature model with suitable boundary conditions. \cite{Ju06,Miranda11} 
From these theoretical studies,\cite{Ju06,Miranda11} the non-equilibrium behavior between 
electrons and atoms occurs at the metal-semiconductor interface, and as the thickness of 
metal layers increase, the electrons and atoms are in better equilibration inside the metal 
layers. 

\begin{figure}[htbp]
\includegraphics[width=5in]{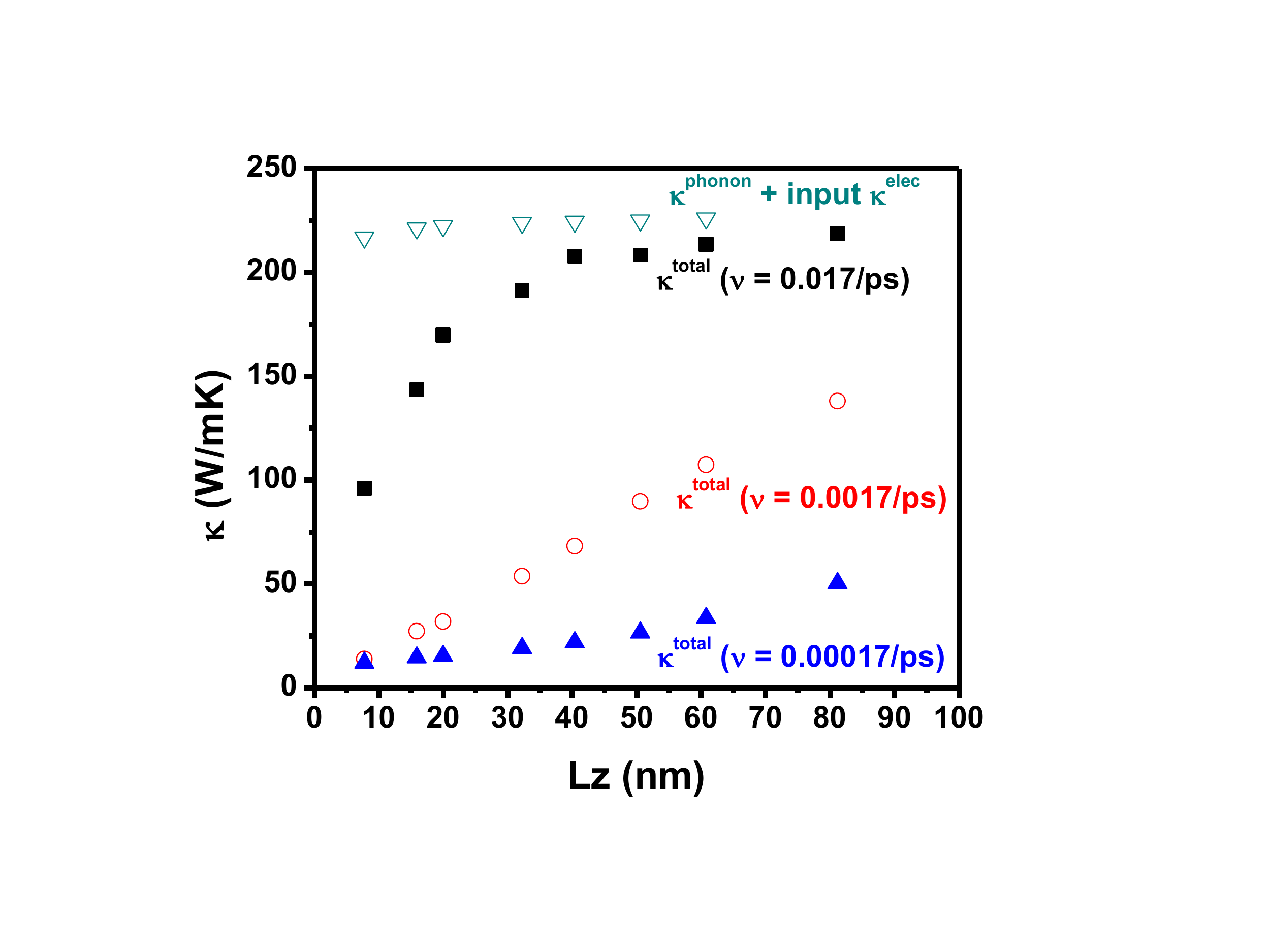}
\caption{ \label{Fig:ThermalCond}
Thermal conductivity as a function of specimen length for various electron-phonon
coupling constants. Green open triangles denote phonon thermal conductivity plus input
electron value.}
\end{figure}

Figure \ref{Fig:ThermalCond} shows the calculated thermal conductivity of the Al specimens
as a function of their size for various coupling constants investigated. The ideal result
of the atomic contribution (from the $\nu$=0.00 ps$^{-1}$ simulations) plus the input electronic thermal conductivity $\kappa^{elec}$ is shown as green open triangles. This value is only reached when electrons and atoms
reach local equilibrium  within the specimen. Even under steady state the heat exchange regions
are driven away from equilibrium and local equilibrium within the sample is only achieved for relatively 
long specimens or large coupling constants. As the specimen size or the electron-phonon coupling rate 
is reduced the lack of local equilibration between the two subsystems leads to electrons contributing sub-optimally 
to thermal transport and a decrease in the effective thermal conductivity of the material. Similar effects
have been observed in purely phononic transport in cases with different phonon groups remain
away from local equilibrium \cite{Hu11,zhou13}.

This example shows the power of coupling a two-temperature description with MD simulations for
ions can naturally capture non-diffusive effects and non-equilibrium processes.

\section{Conclusions and Outlook}

All particle-based simulations of materials describe explicitly only
a limited set of degrees of freedom which are considered essential
for the problem at hand; many other DoFs are treated in an approximate
manner or their contributions disregarded altogether. This approximation
puts severe constraints to the type of phenomena that can be described
accurately by these methods.

In this paper we have presented the implementation of a thermodynamically 
accurate set of mesodynamics equations of motion that accurately describe
the thermal and transport roles of the implicit DoFs. We applied this
method to describe intra-molecular degrees of freedom in a mesoscale 
description of a polymer and to describe valence electrons in atomistic
simulations of metallic systems. The mesodynamics enables for a quantum
mechanical treatment of the thermal role of the implicit DoFs. This
leads, in the case of molecular systems, to a mesoscopic description that 
is more accurate than the computationally more intensive all-atom MD 
(which is always classical).

The new thermo-mechanical formulation of mesodynamics presented in this  
paper is generally applicable, extending the spatial and temporal range  
of more expensive all-atom simulations to thermodynamically realistic  
mesoscopic simulations, with the possibility of solving a wide variety of problems in physics, chemistry, materials science, and biology.

\acknowledgements

This work was supported by the ASC Materials and Physics Modeling Program at Los Alamos National Laboratory and
by the US National Science Foundation, Grant ECCS-1028667.

\vspace{0.2in}

\appendix

\section{Thermostatted equations of motion}

Under time-reversible integral feedback, thermo-mechanical mesodynamics  
can be formulated to give time averages that correspond to ensemble  
averages under isothermal, isochoric (NVT) conditions. Under the quasi-ergodic 
hypothesis, the long-time  
average of a quantity measured along a given many-body trajectory  
equals its average, at a given instant of time, over a multitude  
(ensemble) of such trajectories.

To obtain NVT equations of motion from Eqs. \ref{DID}, we first set  
the coupling between internal DoFs to zero, $\dot{\xi}_i = 0$. Next, we take  
the range of the weighting function to be infinite ($w=1$), so that the  
external temperature of all mesoparticles is $T^{ext}_i = T$, and then  
set the number of internal DoFs to infinity, such that the internal 
(thermostat) temperature is fixed for all mesoparticles, $T^{int}_i = T_0$ (this 
represents the temperature of the heat bath). In that  
case, the heat-flow variable also takes on global character: $\zeta_i =  
\zeta$, and, instead of Eq.\ref{zetadot}, its equation of motion is  
derived from the requirement that the canonical ensemble distribution  
function be a stationary solution of the Liouville equation for the  
phase-space motion. We represent the $(2dN+1)$-dimensional phase space  
of coordinates and velocities by ${\bf x} = \{ {\bf r},{\bf u};\zeta  
\}$; the equations of motion, $\dot{{\bf x}}({\bf x},t) =$

\begin{eqnarray}\label{xdot}
{\bf \dot{r}}_i &=& {\bf u}_i + \frac{\nu \zeta}{m_i \omega_E^2}{\bf  
F}_i\nonumber\\
{\bf \dot{u}}_i &=& \frac{{\bf F}_i}{m_i}\nonumber\\
\dot{\zeta} &=& ?
\end{eqnarray}

\noindent where the equation of motion for $\zeta$ is to be determined  
(see Ref. \onlinecite{holian95} for a comparable treatment of Nos\'e-Hoover  
thermostatting).

The canonical distribution function (equilibrium) is

\begin{equation}\label{distfcn}
\rho_0 = \frac{1}{Q_0}\ e^{-\beta E({\bf x})}\ ,
\end{equation}

\noindent where $Q_0$ is the canonical partition function  
(normalization), $\beta = 1/kT_0$, and the total system energy includes  
$\frac{1}{2}kT_0$ for the thermostat heat-flow variable $\zeta$,  
compared with $O(N)$ for the kinetic and potential contributions:

\begin{equation}\label{ecan}
E({\bf x}) = K(\{{\bf u}\}) + \Phi(\{{\bf r}\}) + \frac{d}{2}NkT_0  
\zeta^2\ .
\end{equation}

The distribution function obeys the Liouville continuity equation in  
the (2d+1) dimensional phase space:

\begin{equation}\label{Liouville}
\frac{\partial \rho}{\partial {\bf t}} + \frac{\partial}{\partial {\bf  
x}} \cdot \left(\rho \dot{{\bf x}}\right) = 0\ .
\end{equation}

\noindent Since the equilibrium canonical distribution function does  
not explicitly depend on time (it is a stationary solution of the  
trajectories that make up the ensemble),

\begin{eqnarray}\label{station}
0 &=& \frac{\partial \rho_0}{\partial t}+ \dot{{\bf x}} \cdot  
\frac{\partial \rho_0}{\partial {\bf x}} + \rho_0  
\frac{\partial}{\partial {\bf x}} \cdot \dot{{\bf x}}\nonumber\\
&=& -\rho_0 \beta \dot{E} + \rho_0 \frac{\nu  
\zeta}{\omega_E^2}\sum_{i=1}^N \frac{1}{m_i}\frac{\partial}{\partial  
{\bf r}_i} \cdot {\bf F}_i\nonumber\\
\Rightarrow \dot{E} &=& -dNkT_0 \nu \zeta \  
\frac{\omega^2}{\omega_E^2}\ ,
\end{eqnarray}
where we imposed the condition that  $\dot{\zeta}$ does not depend on
$\zeta$ itself.

Substituting the equations of motion (Eqs. \ref{xdot}) into  
Eq.\ref{station}, we find that

\begin{eqnarray}\label{ecandot}
\dot{E} &=& \sum_i m_i \dot{{\bf u}}_i \cdot {\bf u}_i + \sum_i  
\frac{\partial \Phi}{\partial {\bf r}_i} \cdot \dot{{\bf r}}_i +  
dNkT_0\ \zeta \dot{\zeta}\nonumber\\
&=& \sum_i {\bf F}_i \cdot {\bf u}_i - \sum_i {\bf F}_i \cdot \left({\bf u}_i + 
\frac{\nu \zeta}{m_i \omega_E^2}{\bf F}_i \right) + dNkT_0\  
\zeta \dot{\zeta}\nonumber\\
&=& -\frac{\nu \zeta}{\omega_E^2}\ \sum_i \frac{|{\bf F}_i|^2}{m_i} +  
dNkT_0\ \zeta \dot{\zeta}\nonumber\\
\Rightarrow \dot{\zeta} &=& \nu \left(\frac{1}{dNkT_0\omega_E^2}\  
\sum_i \frac{|{\bf F}_i|^2}{m_i} - \frac{\omega^2}{\omega_E^2} \right)\  
.\
\end{eqnarray}

\noindent At long times, the equation of motion for the heat-flow  
variable $\zeta$ gives a time average equal to zero:

\begin{eqnarray}\label{long}
\overline{\dot{\zeta}} &=& \lim_{t\rightarrow\infty}\  
\frac{1}{t}\int_0^t\ ds\ \frac{d\zeta}{ds} = \lim_{t\rightarrow\infty}\  
\frac{\zeta(t)-\zeta(0)}{t} = 0 \nonumber\\
&\Rightarrow& \overline{\omega^2} = \frac{1}{dNkT_0}\ \sum_i  
\frac{\overline{|{\bf F}_i|^2}}{m_i}\ .
\end{eqnarray}

We can evaluate the canonical ensemble average of the mean-square  
frequency as follows:

\begin{eqnarray}\label{canom1}
\langle \omega^2 \rangle &=& \frac{1}{Q_0}\prod_j\int_V d{\bf r}_j  
e^{-\beta \Phi}\ \frac{1}{dN}\ \sum_i  
\frac{1}{m_i}\frac{\partial}{\partial {\bf r}_i} \cdot \frac{\partial  
\Phi}{\partial {\bf r}_i}\nonumber\\
&=& \frac{1}{d} \int_V d{\bf r} e^{-\beta \Phi}  
\frac{1}{m}\frac{\partial}{\partial {\bf r}} \cdot \frac{\partial  
\Phi}{\partial {\bf r}}\ {\Big /} \int_V d{\bf r} e^{-\beta \Phi}\ ,
\end{eqnarray}

\noindent where integration by parts (with the surface term equal to  
zero under periodic boundary conditions) yields

\begin{eqnarray}\label{parts}
\int_V d{\bf r} e^{-\beta \Phi} \frac{1}{m} \frac{\partial}{\partial  
{\bf r}} \cdot \frac{\partial \Phi}{\partial {\bf r}} &=& \frac{1}{m}  
\frac{\partial \Phi}{\partial {\bf r}} \cdot \hat{{\bf n}}_S e^{-\beta  
\Phi} {\Big |}_{-S}^{+S}\nonumber\\
&&- \ \int_V d{\bf r} \frac{1}{m} \frac{\partial \Phi}{\partial {\bf  
r}} \cdot \frac{\partial}{\partial {\bf r}} e^{-\beta \Phi}\nonumber\\
&=& \beta \int_V d{\bf r} \frac{1}{m} \frac{\partial \Phi}{\partial  
{\bf r}} \cdot \frac{\partial \Phi}{\partial {\bf r}} e^{-\beta  
\Phi}\nonumber\\
&=& \beta \int_V d{\bf r} e^{-\beta \Phi} \frac{|{\bf F}|^2}{m}\ ,
\end{eqnarray}

\noindent so that

\begin{equation}\label{canom2}
\langle \omega^2 \rangle = \frac{1}{dNkT_0}\sum_i \frac{\langle |{\bf  
F}_i|^2 \rangle}{m_i}\ ,
\end{equation}
this relation between $\langle \omega^2 \rangle$, the temperature and
force squared can also be obtained from the velocity power spectrum.

\noindent We see, therefore, that mesodynamics thermalization can be  
specialized to global canonical thermostatting, where the equations of  
motion satisfy the quasiergodic hypothesis: $\omega_E^2 =  
\overline{\omega^2} = \langle \omega^2 \rangle$.

 From these results, it is tempting to postulate a new definition of  
instantaneous temperature $\tilde{T}$, written in such a way as to  
easily identify mass times velocity-squared (see the definition of the  
dissipative terminal velocity in Eq.\ref{terminal}),

\begin{equation}\label{tempf}
dk\tilde{T} = \frac{1}{N}\sum_i m_i {\Big|}\frac{{\bf F}_i}{m_i  
\omega_E}{\Big|}^2\ ,
\end{equation}

\noindent by analogy with the usual kinetic definition $T$ (see, e.g.,  
Eq.\ref{exttemp}, with the weighting function set to $w \equiv 1$),

\begin{equation}\label{tempv}
dkT = \frac{1}{N}\sum_i m_i |{\bf u}_i - \langle {\bf u} \rangle|^2\ .
\end{equation}

\noindent $\langle {\bf u} \rangle$ is the c.m.\ velocity of the system  
(see e.g., Eq.\ref{extvel}). However, defining the temperature in terms  
of forces differs instantaneously from that defined by velocities, as  
can be exemplified most clearly by hard spheres: most of the time, the  
forces are zero as particles move in straight-line, constant-velocity  
trajectories between collisions; at the moment of collisions, forces  
are formally infinite and velocities are zero; therefore, the two  
definitions of ``instantaneous'' temperature are starkly different,  
though their running time-averages do approach each other, as can be  
seen from Eq.\ref{ecandot}, into which we substitute the new  
temperature definition of Eq.\ref{tempf}:

\begin{equation}\label{zetadotNH}
\dot{\zeta} = \nu \left(\frac{\tilde{T}}{T_0} -  
\frac{\omega^2}{\omega_E^2}\right)\ ,
\end{equation}

\noindent and where the long-time average gives $\overline{\tilde{T}} =  
T_0$.

Notice that this mesodynamics approach to the canonical ensemble does  
not at all replace the Nos\'e-Hoover (NH) approach to bulk  
thermostatting \cite{NH}, though Eq.\ref{zetadotNH} bears a strong  
resemblance to the latter's equation of motion for the heat-flow  
variable. For one thing, the mesodynamics version requires an expensive  
evaluation of the Hessian (curvature) of the potential energy;  
moreover, two force evaluations per central-difference time-step are  
required: one for the coordinate update, and one for the velocity  
update. Thus, the traditional NH thermostat is at least twice as  
computationally efficient as the one derived from mesodynamics.

While the derivation of the canonical thermostat from mesodynamics that  
we have presented here is only of academic interest, it shows  
nevertheless, that the thermo-mechanical foundations of mesodynamics  
are deep. On the other hand, the need for twice the number of force  
evaluations is simply the price that has to be paid for Galilean  
invariance of the equations of motion.

\section{Finite central-difference approximation to the  
mesodynamics equations of motion}

In this Appendix, we display the finite central-difference equations  
(originally the St{\o}rmer method, which was ``rediscovered'' by  
Vineyard \cite{vineyard}---but also known commonly in the modern  
literature as the ``Verlet'' method); the computational time-step is  
$\delta$. In general, coordinates are computed at integer time-steps,  
while velocities are computed at half-integer values. In common  
parlance, this is called the ``leap-frog'' method, since, in the update  
of either coordinate or velocity, the time derivative is evaluated  
halfway in between the new and old time-steps. The time derivatives are  
always computed at least to $O(\delta^2)$, which when multiplied by  
$\delta$ give errors in the Taylor series that are formally of  
$O(\delta^3)$. When the updates of velocities and coordinates, via the  
first-order ordinary differential equations (o.d.e.'s), are combined  
into second-order o.d.e.'s for coordinates alone, the local error in  
the finite central-difference equations are $O(\delta^4$); central  
differences are referred to as a second-order integration method.

In this paper, the mesodynamics equations of motion are presented in  
two flavors: (1) integral feedback in Eqs.\ref{SHEOM_integral} and (2)  
direct feedback in Eqs.\ref{SHEOM_direct}. (For simplicity in the  
following development, we drop the vector notation for coordinates,  
velocities, and forces, as well as the mesoparticle index $i$.) The  
equation of motion for the energy of the internal degrees of freedom  
can be easily generalized for a heat capacity that depends on internal  
temperature as a power law, namely, $C_V = CT^n$ (the most interesting  
cases are $n = 0$, where $C_V$ is independent of $T$, and $n = 1$,  
which is appropriate for low-temperature metals); hence, $\dot{E}^{int}  
= (n+1)C_V \dot{T}^{int}$. Therefore, we set $C_V^{\prime} = (n+1)C_V$  
and $\lambda^{\prime} = \lambda /(n+1)$, where $\lambda = \kappa / C_V$  
is the thermal diffusivity.

In the integral feedback version, the mesodynamics difference equations  
[formally to $O(\delta^3)$] are given by:

\begin{eqnarray}\label{SHEOM_finite}
r(t) &=& r(t-\delta) +  
\chi(t-\frac{\delta}{2})F(t-\frac{\delta}{2})\delta \nonumber\\
u(t+\frac{\delta}{2}) &=& u(t-\frac{\delta}{2}) + \frac{F(t)}{m} \delta  
\nonumber\\
T^{int}(t+\frac{\delta}{2}) &=& T^{int}(t-\frac{\delta}{2}) +  
\frac{\chi(t)|F(t)|^2}{C_V^{\prime}} \delta \nonumber\\
&&+ \ \frac{\nu_0 \xi(t)}{C_V^{\prime}} \delta \nonumber\\
\zeta(t+\delta) &=& \zeta(t) + \nu \frac{T^{ext}(t+\frac{\delta}{2}) -  
T^{int}(t+\frac{\delta}{2})}{T_0} \delta \nonumber\\
\xi(t+\delta) &=& \xi(t) + \kappa \nabla^2 T^{int}(t+\frac{\delta}{2})  
\delta \ ,
\end{eqnarray}

\noindent where $\chi = \nu \zeta/m \omega_E^2$ and $T_0$ is an  
arbitrary temperature (the initial value of $T^{int}$, for example).  
The auxiliary flow variables $\zeta$ and $\xi$ are most naturally  
evaluated at integer time-steps; whenever they appear with half-integer  
values of time in the above equations, they are being used in mid-point  
time derivatives, so that they can be evaluated as averages over the  
neighboring integer time-steps: for example,

\begin{equation}\label{halftime}
\zeta(t+\frac{\delta}{2})=\frac{1}{2}\left[\zeta(t)+\zeta(t+\delta)\right] + O(\delta^2)\ .
\end{equation}

In the coordinate update (the first line of Eq.\ref{SHEOM_finite}), the  
force that appears must be evaluated at the end of the previous  
time-integration cycle---the temporary coordinates used to compute the  
forces are evaluated to $O(\delta^2)$ and stored as temporary  
values---thereupon, the force is evaluated and stored for the next  
time-step. Thus, the force needs to be evaluated twice in each  
time-step cycle, rather than just once, as is usual in standard  
molecular-dynamics simulations.

The direct feedback version of the central-difference equations is  
given by

\begin{eqnarray}\label{SHEOM_direct}
r(t) &=& r(t-\delta) + \chi(t-\frac{\delta}{2})F(t-\frac{\delta}{2})  
\delta \nonumber\\
u(t+\frac{\delta}{2}) &=& u(t-\frac{\delta}{2})) + \frac{F(t)}{m}  
\delta \nonumber\\
T^{int}(t+\frac{\delta}{2}) &=& T^{int}(t-\frac{\delta}{2}) +  
\frac{\chi(t)|F(t)|^2}{C_V^{\prime}} \delta \nonumber\\
&&+ \ \lambda^{\prime} \nabla^2 T^{int}(t) \delta \ ,
\end{eqnarray}

\noindent where $\chi = \nu (T^{ext} - T^{int})/T_0$. Note that the heat conduction term (Fourier's Law) needs an estimate (a temporary variable) for $T^{int}(t)$ that is accurate to $O(\delta^2)$:

\begin{eqnarray}\label{test}
T^{int}(t) &=& T^{int}(t-\frac{\delta}{2}) + \frac{\chi(t)|F(t)|^2}{C^{\prime}} \frac{\delta}{2} \nonumber\\ 
&&+ \ \lambda^{\prime} \nabla^2 T^{int}(t-\frac{\delta}{2}) \frac{\delta}{2} \ ,
\end{eqnarray}

\noindent from which $\nabla^2 T^{int}(t)$ is then computed for use in Eq.\ref{DID_direct}.

\end{document}